\documentclass[10pt,a4paper,notitlepage]{article}
\usepackage{geometry}             
\usepackage{extsizes}
\usepackage{amssymb}
\usepackage{amsmath}
\usepackage{amsthm}
\usepackage{url}
\usepackage{placeins}

\usepackage{subfigure}
\usepackage{multirow}

\usepackage{epsfig}

\newcommand{\E}{\mathop{\mathbb{E}}{}\!}

\newcommand{\Q}{\ensuremath{\mathbb{Q}}{}\!}

\newcommand{\Gp}{\ensuremath{\mathcal{G}}}
\newcommand{\Pp}{\ensuremath{\mathcal{P}}}
\newcommand{\Dp}{\ensuremath{\mathcal{D}}}

\newcommand\be{$$}
\newcommand\ee{$$}
\newcommand\ben{\begin{equation}}
\newcommand\een{\end{equation}}
\newcommand\bea{\begin{eqnarray*}}
\newcommand\eea{\end{eqnarray*}}
\newcommand\bean{\begin{eqnarray}}
\newcommand\eean{\end{eqnarray}}

\newcommand\h{\ensuremath{\frac{1}{2}}}

\newcommand\Vh{\ensuremath{\widehat{V}}}
\newcommand\cA{\ensuremath{\mathcal{A}}}

\newcommand\GW{\ensuremath{{G}}}

\newcommand\AM{\ensuremath{{\rm AM}}}

\begin{document}
\author{Chris Kenyon and Richard Kenyon\footnote{Contact: chris.kenyon@lloydsbanking.com; Richard D. Kenyon is a Senior Lecturer in Accountancy at Northampton Business School}}
\title{DVA for Assets \footnote{\bf The views expressed are those of the authors only, no other representation should be attributed.}}
\date{08 January 2013, Version 1.7\\ \vskip3mm {\it A version of this paper will appear\\ in Risk in February 2013}}

\maketitle

\begin{abstract}
The effect of self-default on the valuation of liabilities and derivatives (DVA) has been widely discussed but the effect on assets has not received similar attention.  Any asset whose value depends on the status, or existence, of the firm will have a DVA.  We extend \cite{Burgard2011a} to provide a hedging strategy for such assets and provide an in-depth example from the balance sheet (Goodwill).  We calibrate our model to seven US banks over the crisis period of mid-2007 to 2011.  This suggests that their reported profits would have changed significantly if DVA on assets, as well as liabilities, was included --- unless the DVA was hedged.
\end{abstract}

\section{Introduction}

The effect of self-default on the valuation of liabilities and derivatives (DVA) has been widely discussed in the pricing literature \cite{Burgard2011a, Brigo2011a, 
Cesari2010a, Pallavicini2011b, Crepey2012b}.  However, the effect of self-default on assets has yet to attract similar attention (\cite{Kenyon2012a} being an exception), although it is clear that default will affect any asset that depends on company existence or performance.  

We provide a hedging strategy for pricing DVA on assets extending \cite{Burgard2011a}, and consider an example, Goodwill, in depth.  We calibrate our model to seven US banks over the crisis period of mid-2007 to 2011 and show how their reported profits would have changed if DVA on this asset, as well as liabilities, had been included.  This effect is highly significant for at least four of the seven banks.

FAS 157 requires US banks to reflect their own potential nonperformance, which includes creditworthiness, in the fair value of their liabilities \cite{FASB-157}.  However, creditworthiness has effects on balance sheet items beyond liabilities. This can be observed by their change in value upon default of the company holding them.  This may appear surprising, but it is clear that any asset that relies on the company being a going concern will exhibit this behavior, e.g. Goodwill, brand values, etc.  In fact Goodwill can be written down prior to default, and thus have a major effect on balance sheets even for going concerns.  We include this in our model and calibration.  Thus we demonstrate how FAS 157 can be applied to the asset side of the balance sheet as well as the liability side.  We do not propose a change in how Goodwill is derived \cite{Ramanna2010a}.  Instead we propose an adjustment that is applied subsequently to reflect creditworthiness effects.

\section{Hedging DVA on Assets\label{s:hedging}}

We take the view that own-assets can be sensitive to own-stock price levels as well as own-default.  Thus we can model, for example, progressive writedowns on a banks' Goodwill as its stock price decreases.  We modify  \cite{Burgard2011a} in that we have no risky counterparty, and extend it in that the own-asset (the bank stock) $S(t)$ jumps to zero on bank default.   As \cite{Burgard2011a} we assume that a riskless bond can be purchased.  Thus under the historical measure we have:
\bean
dP(t) / P(t-) &=& r dt \nonumber \\
dP_b(t) / P_b(t-) &=& r_b dt - dJ_b  \nonumber \\
dS(t) / S(t-) &=& \mu dt + \sigma dW - d J_b \label{e:stock}
\eean
where:

$P,\ P_b(t)$: price of riskless and risky bonds respectively;

$r,\ r_b$: riskless interest rate and risky rate applicable to purchased and issued bonds respectively (issued bonds can be repurchased);

$W$: Brownian driving process;

$J_b$: jump to default of the bank;

$S$: is the stock of the bank.  Note that the only jump in $S$-value comes on bank default.  There are no market-based jumps in $S$-value.

\noindent We could use a non-zero recovery on the bank's issued bonds, but we assume zero recovery for computational convenience as in \cite{Burgard2011a}.  

Let \Vh\ be the value of an own-asset that depends on the bank's own stock, and the banks existence (i.e. it also depends on bank default).  If the bank defaults at $\tau$ then:
\be
\Vh(\tau,S)= M^+(\tau,S) + R_b M^-(\tau,S)
\ee
where $M$ is the value of the own-asset at default.  We keep this value general for now (allowing positive and negative values).  This enables us to model either hedging the asset or hedging the loss on the asset on default, which will be important later.    

Our setup is simpler than \cite{Burgard2011a} in that we only need to consider own-default, however we include a risky underlying ($S(t)$) which has consequences.  The value $V^\Pi(t)$ of the hedging portfolio $\Pi(t)$ can be written in terms of the price processes $\Pp_*$ of its components:
\be
-V^\Pi(t) =\Pi(t) = \delta(t) \Pp_S(t) + \alpha_b(t)\Pp_{P_b}(t) + \Pp_\beta(t)
\ee
where $\delta(t)$ is the quantity of stock held, $\alpha_b(t)$ risky bond holdings, and $\Pp_\beta(t)$ is the price of the cash.  We require the portfolio to be self-financing, so (bearing in mind \cite{Brigo2012a}) we have the following gain $\Gp_*$ processes:
\bea
d\Gp_S &=& dS + (\gamma - q)S dt \\
d\Gp_{P_b} &=& r_b P_b dt - P_b dJ_b\\
d\Gp_\beta &=&  r \epsilon^+dt + r_F \epsilon^-  dt
\eea
Note that all gain processes are functions of $t-$ not $t$.  $\gamma_S(t)$ is the dividend yield on $S(t)$ and $q_S(t)$ is the financing cost.  As in \cite{Burgard2011a}, we assume we can put $S(t)$ into repo and we also assume it closes flat on default.  Equally we assume zero recovery for the stock lender when we are short selling.  If the cash position is positive, riskless investment yields $r$, whereas negative cash costs the funding rate $r_F$.  We can set the funding rate to the yield of an issued bond with recovery $R_b$, so $r_F=r+(1-R_b)\lambda_b$.    The price processes $\Pp_*$ are:
\be
\Pp_S = 0;\qquad \Pp_{P_b} = P_b;\qquad \Pp_\beta = \epsilon.
\ee
The stock price process is zero except exactly at the instant of default but this portfolio cannot be bought (no trading exactly at $\tau$).  Note that the dividend processes $\Dp_*$ (with time zero values of zero) are not individually zero:
\bea
d\Dp_S     &=& d\Gp_S - d\Pp_S         = dS + (\gamma - q)S dt \\
d\Dp_{P_b} &=& d\Gp_{P_b} - d\Pp_{P_b} = r_b P_b dt - d P_b  \\
d\Dp_\beta &=& d\Gp_\beta - d\Pp_\beta = r \epsilon^+dt + r_F \epsilon^- dt 
\eea
Self-financing requires that $\Gp^\Pi=V^\Pi$ and replication requires that $V^\Pi=\Vh$, so $G^\Pi = \Vh$.  Considering $V^\Pi$ we have:
\be
-V^\Pi = \delta \times 0 + \alpha_b \times P_b + 1 \times \epsilon
\quad \implies \quad \epsilon = -\Vh - \alpha_b P_b
\ee
Now the portfolio gain process $\Gp^\Pi$ is by definition the weighted sum of the individual gains, hence:
\bean
d\Gp^\Pi &=& \delta(dS + (\gamma - q)S dt) + \alpha_b (r_b P_b dt - P_b dJ_b)\nonumber \\
&&{}+ \{ r (-\Vh - \alpha_b P_b)^+ + r_F (-\Vh - \alpha_b P_b)^-   \} dt
\label{e:dG}
\eean
Applying Ito's lemma to \Vh\ we have:
\bean
d\Vh &=& \partial_t \Vh dt + \partial_S \Vh dS + \h \sigma^2 S^2\partial_{SS} \Vh dt 
\nonumber \\
&&{}+ (\Vh(\tau) -\Vh(\tau-) - S(\tau-)\partial_S\Vh(\tau-)) dJ_b
\label{e:Ito}
\eean
where $\tau$ is the default time of the bank and that the bank can only jump once to default.  Notice that the jump term in $dS$ leads to one additional term within the last bracket.  Positive \Vh\ means long $b$-risk so the coefficient of $P_b$, $\alpha_b$ will be positive or zero.  

Removing all risks by equating the $dS$ and $dJ_b$ coefficients within Equations \ref{e:dG} and \ref{e:Ito} (which are equal):
\bea
\delta(t) &=& \partial_S \Vh(t) \\
\alpha_b(t) &=&- (\Vh(\tau) -\Vh(\tau-) - S(\tau-)\partial_S\Vh(\tau-))/P_b(\tau-) 
\eea
If we now define a parabolic differential operator $\cA_t$ as:
\ben
\cA_t<> := \h \sigma^2 S^2 \partial_{SS}<> + \left\{\lambda_b+(q_S - \gamma_S)\right\}S\partial_S<>  \label{eq:op}
\een
and
\bea
\lambda_b &:=& r_b - r \\
s_F &:=& r_F - r 
\eea
then \Vh\ satisfies the PDE
\bean
\partial_t\Vh + \cA_t \Vh - r \Vh 
&=& s_F(\Vh + \Delta \Vh +S\partial_S\Vh )^+ - \lambda_b (\Delta \Vh + S\partial_S\Vh)  \label{e:pde2} \\
&=& \lambda_b \Vh + s_F (M^+ + R_bM^- + S\partial_S\Vh)^+ \nonumber\\
&&{}-\lambda_b(R_bM^-+M^+ + S\partial_S\Vh)  \label{e:pde3} \\
\partial_t\Vh + \widetilde{\cA_t} \Vh - r \Vh 
&=&
\lambda_b \Vh + s_F (M^+ + R_bM^- + S\partial_S\Vh)^+ \nonumber\\
&&{}-\lambda_b(R_bM^-+M^+)  \label{e:pde4} 
\eean
We move to Equation \ref{e:pde4}  by absorbing the $\lambda_b S\partial_S\Vh$ term into $\cA_t$ relabelling it $\widetilde{\cA_t}$.  Note that the terms inside the $s_F$-bracket are evaluated with the share  price pre-default (funding occurs only whilst not defaulted), whereas the terms inside the $\lambda_b$-bracket are evaluated post-default.

We have been able to remove the jump risk from the PDE because the value of the stock is known pre-default and its jump size on default is also known.  We needed to use both the stock itself and the own-bond to remove all the jump risk from the portfolio (especially the stock), i.e. both $\delta$ and $\alpha_b$ were used.

Below we assume that we can put \Vh\ into repo so $s_F=0$ hence:
\ben
\partial_t\Vh + \widetilde{\cA_t} \Vh - r \Vh = \lambda_b(\Vh -(R_bM^-+M^+) )  \label{e:pde5}
\een

\section{Example Asset: Goodwill\label{s:DVAonGW}}

\subsection{Introduction to the Asset}

\begin{figure}[htbp]  
\begin{minipage}[b]{0.5\linewidth}  
\centering 
\includegraphics[width=1.05\textwidth]{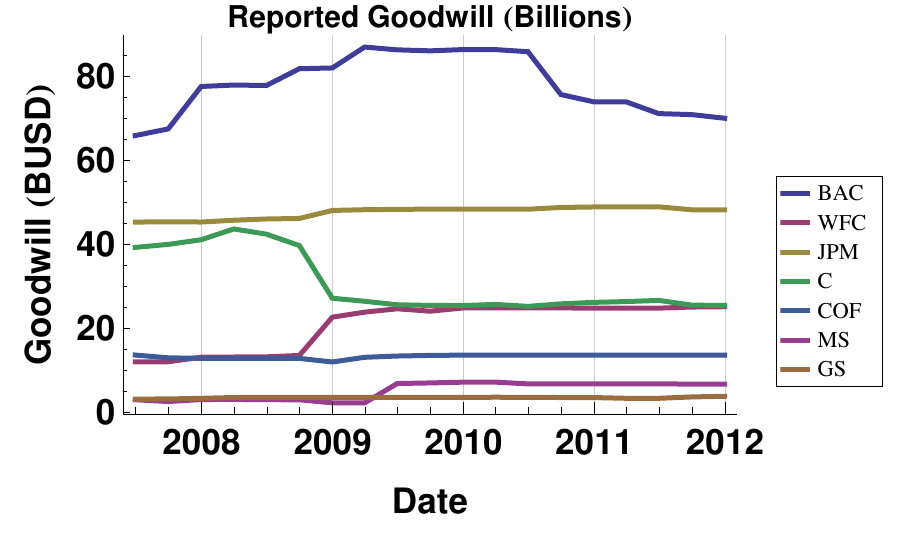}  
\end{minipage}  
\hspace{0.5cm}  
\begin{minipage}[b]{0.5\linewidth}  
\centering 
\includegraphics[width=\textwidth,trim=15 0 0 0]{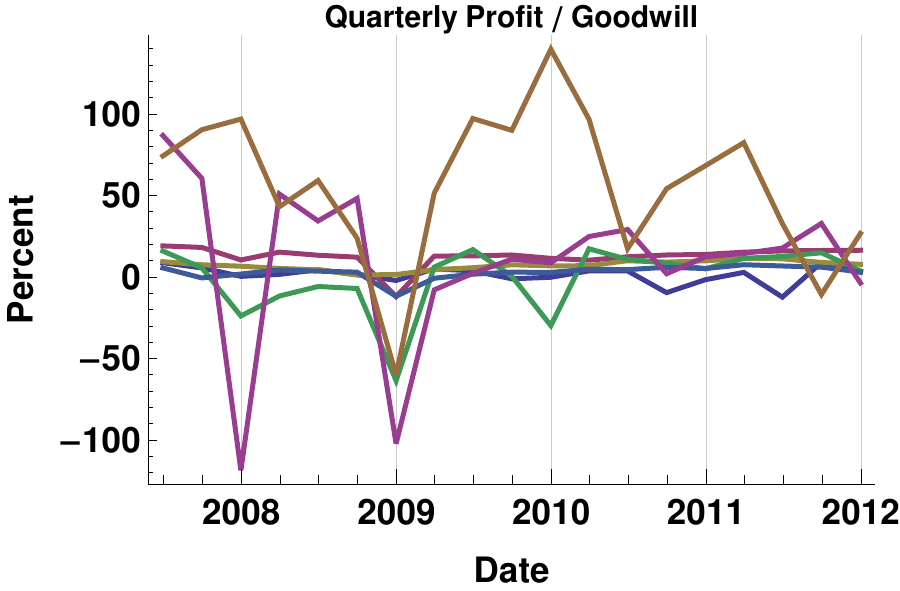}  
\end{minipage} 
\caption{{\bf Left panel:} Reported Goodwill for seven US banks (Bloomberg tickers: BAC; WFC; JPM; C; COF; MS; GS) in Billion USD. {\bf Right panel:} Quarterly profits as a percentage of Goodwill.  For the non-pure-investment-banks this is a small percentage.  For the two investment banks (GS, MS) Goodwill is relatively small and profits are the same order of magnitude. Data is from Bloomberg.} 
\label{f:ProfitGW} 
\end{figure} 

Goodwill is an asset on the balance sheet that is reported quarterly.  In our examples  we consider seven large US banks, five are chosen for size and the other two as representative pure investment banks.  These banks have significant levels of Goodwill on their balance sheets as compared to quarterly profits, see Figure \ref{f:ProfitGW}.  

Goodwill (defined in FAS 350-20) is created, amongst other possibilities, when a company is bought for more than the fair value of net assets acquired and liabilities assumed.  The carrying value of Goodwill must be regularly reviewed (at least annually, in a two step proceedure\footnote{The first step is to see whether it is more likely than not($>50\%$ likely) that the carrying value exceeds the fair value (FASB updates 2011-08, 2012-02).}), and can only stay at the same level or be impaired, i.e. decrease (FAS 350-20-35-13).  The basis for the carrying value is as though it were purchased anew (-35-14).  Thus we assume that the value of Goodwill is capped on the balance sheet at its initial amount.  

In regulatory terms Goodwill is not counted towards capital (Basel III), effectively saying that it has zero recovery value.  That is, it provides no buffer against default.   IFRS 9 puts it into Other Comprehensive Income, which feeds into Equity, and it will remain on the balance sheet.   

For tax purposes some information (e.g. USC Title 26 A.1.B.VI Section 197) suggests that Goodwill must be amortized over 15 years.  We use the word suggests to remind readers that this is a paper exploring the application of DVA to assets, not a definitive tax or accounting opinion.  This amortization creates tax credits that can reduce tax payments on future profits.  

Tax treatments change present and future cashflows, whereas accounting treatments change reports and opinions.

\FloatBarrier
\subsection{Models for Goodwill}

\begin{figure}[htbp]  
\begin{minipage}[b]{0.5\linewidth}  
\centering 
\includegraphics[width=0.9\textwidth]{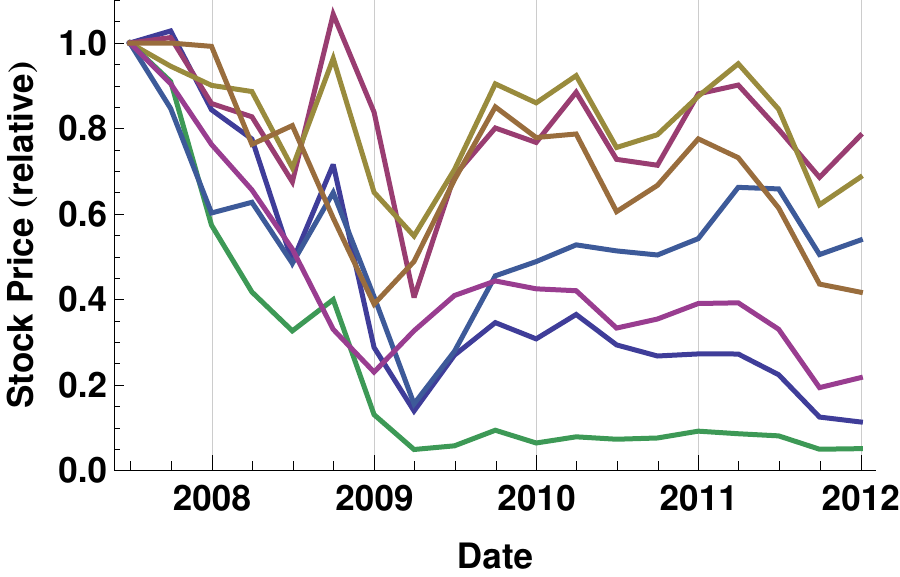}  
\end{minipage}  
\hspace{0.5cm}  
\begin{minipage}[b]{0.5\linewidth}  
\centering 
\includegraphics[width=\textwidth]{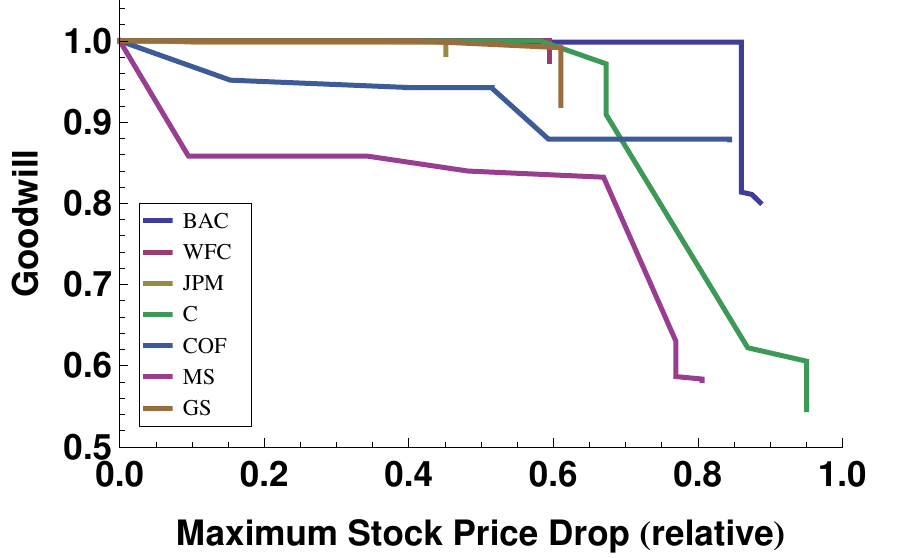}  
\end{minipage} 
\caption{{\bf Left panel:} Stock prices for seven US banks relative to their 2H2007 values. {\bf Right panel:} Derived calibration curves for Goodwill write-offs versus minimum stock price observed since mid 2007.} 
\label{f:EQandGW} 
\end{figure} 

We consider three models of Goodwill value, two inspired by accounting and one from tax.  First examine Figure \ref{f:EQandGW} where we show relative equity prices from mid 2007 on the left, and relative Goodwill value (considering only writedowns)  versus maximum stock price drop on the right.  Despite significant stock price drops three of the seven banks considered wrote down less than 10\%\ of their Goodwill.  Thus we propose the following three models.
\begin{description}
\item[CONSTANT] the dollar value of Goodwill never changes.
\item[PROGRESSIVE] as the stock price drops Goodwill is gradually written off.
\item[AMORTIZING] Goodwill is amortized linearly over a fixed number of years.
\end{description}
The last model is inspired by USC Title 26 A.1.B.VI Section 197, mentioned above.  In general we would use a combination of CONSTANT and PROGRESSIVE since, for example, Bank C did not write down 50\%\ of Goodwill despite a 90\%-plus drop in stock price.

\FloatBarrier
\subsection{DVA on Goodwill}

We define DVA on Goodwill as the expected loss from default or degradation of the company.  Degradation is measured by decline in stock price.  Our hedging methodology above includes both possibilities.  Typically there is no recovery on Goodwill as it represents part of going concern value.  

Goodwill is not a tradeable asset in the normal sense used in mathematical finance.  As a thought experiment consider the simplified case that Goodwill has a constant dollar value up to the default of the company.  No self-financing portfolio could duplicate this value as it always decreases in a risk neutral measure (with positive interest rates) as we consider longer horizons.  Furthermore, Goodwill generates no cashflows (even when written down), but has a non-zero value today.  
 
Although Goodwill itself is not a tradeable asset, DVA on Goodwill is a tradeable asset, at least in the sense that we can hedge it.  We will be precise later, but for now we continue the thought experiment.  Suppose we buy a zero-recovery CDS with notional equal to the Goodwill, very long maturity (sufficiently high that the probability of default is close to unity), zero upfront cost, and a given periodic premium payment.  We can hedge this using a zero-recovery bond from the company, and riskless bank account (ignoring different borrowing and lending costs for the moment, that we cover below) \cite{Carr2005a}.  This CDS perfectly compensates us for the loss of value on default of the company (assuming no counterparty risk on the CDS itself).  This works because unit notional CDS hedge bond notionals not their coupons.  This is the intuition behind the application of the hedging strategy of the Section \ref{s:hedging}.

\subsubsection{CONSTANT Model} 

To hedge the loss of Goodwill value $k$ on default the bank enters into a trade $\Vh$ that pays $k$ on bank default, and zero otherwise.  Note that the hedge will not require a position in $S$.  For $\Vh$ to be riskless we assume that it is collateralized and since the value will always be positive to the bank this means, effectively, that the derivative can be placed into repo and thus $s_F=0$.  Hence $M^+=k$, $M^-=0$ and Equation \ref{e:pde3} becomes:
\be
\partial_t\Vh + \widetilde{\cA_t} \Vh - r \Vh = \lambda_b\Vh - \lambda_b k
\ee
Feynman-Kac transforms this to:
\bean
\Vh &=& {\E^\Q} \left[ \int_0^T e^{-(r+\lambda_b)s}\lambda_b k\ ds\right] \nonumber \\
&=& \frac{\lambda_b}{r+\lambda_b}k-\varepsilon(T) \label{e:const}
\eean
where $\varepsilon(T)$ can be made arbitrarily small as $T\rightarrow\infty$.

Equation \ref{e:const} says that if the hazard rate is large relative to the riskless rate then the DVA on Goodwill will tend to the value of Goodwill itself.  This makes sense because Goodwill is lost on default and a high hazard rate implies this happens soon.  

\subsubsection{AMORTIZING Model of Goodwill}

If Goodwill is amortizing in a straight line this means that, in tax terms, it creates a loss every year that can be offset against profits, these are called tax credits.   We model the future value of the tax credits as:
\be
\GW_{\AM}(t) = \sum_{i=\lceil t \rceil}^{n_A} e^{-r (i - \lceil t \rceil)} \frac{G(0)}{n_A}
\ee
where we have assumed that tax is paid once per year at the end of the year and amortization is over $n_A$ years (an integer).  This is a simple extension of the CONSTANT model because the values are deterministic. Thus, as before the hedge will not require a position in $S$.

In this view of the value of Goodwill there is no link to the value of the company, or to any revaluation of Goodwill after its creation.  The value is set by law.  Of course the DVA on Goodwill is still set by the chance of default, and the relevant hedging strategy.

The value of a tax credit obviously depends on having profits, however the profit does not always have to be in the same year as the tax credit.  There is usually a (limited) ability to move these through time when a company does not have sufficient profits to use up tax credit, which then becomes a deferred tax credit.  The formula above assumes that sufficient profits exist to use up all the tax credit as they appear.  We show below that the DVA in this AMORTIZING model is relatively insensitive to the length of the amortization.  Thus whether the generated tax credits are used immediately, or not, is not highly significant.

\subsubsection{PROGRESSIVE Model}

Goodwill is written off progressively as the equity price declines, considering the minimum stock price reached.  We want to hedge these losses, i.e. this DVA on Goodwill.  

\paragraph{Calibration} In general the precise relationship between equity price declines and Goodwill writeoff is one for internal analysts to answer as they are the ones calculating the value of Goodwill.  However, looking back historically we can recover the calibration curves that internal analysts would have calculated.  

Figure \ref{f:EQandGW} shows the monotonic decline in Goodwill with minimum stock price reached (at quarter ends).  Note that this model leaves open the possibility that the writedown occurs with a delay after the barrier (stock price minimum) is reached.  This can occur, but there is typically just one big drop in the period, so we leave this detail out of the modeling, it is a straightforward extension (delayed cashflow).  The calibration set of barriers and losses is shown in Table \ref{t:GWcalibration}.

\paragraph{Hedging} consists of instruments that give positive cash-flows when the stock price reaches successive barriers.  Thus we can write DVA on Goodwill in terms of a series of American-style binary (cash-or-nothing) options $\Vh_i$.  Practically we have captured aspects of a structural model of approach-to-default with these barriers (as opposed to reduced form).

The hedges are bought options and hence always positive-valued.  Since we do not want counterparty risk on the options we assume that they are collateralized and hence $s_F=0$.  Unlike the CONSTANT case, on default these options pay the loss amounts $l_i$ since the stock price will have breached the respective barrier.  Thus we have a similar PDE but with different boundary conditions:
\ben
\partial_t\Vh_i + \widetilde{\cA_t} \Vh_i - r \Vh_i = \lambda_b\Vh_i - \lambda_b l_i \label{e:prog}
\een
provided $S\ge b_i$ where $(b_i,l_i)$ is the (barrier,loss) pair.

We can evaluate Equation \ref{e:prog} as an integral over standard one-touch options with rebates for not touching with different maturities, since default is independent of stock price. A one-touch option is not a free boundary problem so the PDE we have can be used with suitable boundary coniditions, see \cite{Wilmott2006a1} 9.7 for details.
We choose the rebates to be $l_i \lambda_b e^{-(r+\lambda_b)T}$ for maturity $T$, thus capturing the payout at default when the stock price goes to zero, provided the barrier has not previously been reached.  

The no-hit-rebate-at-$T$-of-$R$ option price ${\rm RnoH}(T,R)$ is known:
\be
{\rm RnoH}(T,R) = Re^{-rT} - R\left((S/b_i)^{2\zeta} P_d(b_i^2/S;b_i) + C_d(S;b_i) \right)
\ee
where $C_d,\ P_d$ are digital call and put options and $\zeta=1/2 - (r - \gamma)/\sigma^2$.  
Hence the DVA on Goodwill under the PRGRESSIVE model is:
\be
\sum_i\Vh_i = \sum_i\int_0^T  l_i  \left( {\rm OneTouch}(b_i,s) + {\rm RnoH}(s,1) \right) \lambda_b e^{-\lambda_b s}  \ ds
\ee

It is beyond the scope of this paper to --- from the outside --- estimate future calibrations for Goodwill.  However, making the assumption that internal teams could create a good calibration, we can use the historical data to reproduce it after the fact, see Figure \ref{f:EQandGW} and Table \ref{t:GWcalibration}.  Thus we can calculate their DVA on Goodwill throughout the crisis period as would have been reported.

\begin{table}[htb]
	\centering
		\begin{tabular}{rr| rr| rr| rr| rr| rr| rr}
			\multicolumn{2}{c}{BAC} & \multicolumn{2}{c}{WFC} & \multicolumn{2}{c}{JPM}
			& \multicolumn{2}{c}{C} & \multicolumn{2}{c}{COF} & \multicolumn{2}{c}{MS}
			& \multicolumn{2}{c}{GS} \\
			$b$ & loss & $b$ & loss & $b$  & loss 
			& $b$ & loss & $b$ & loss & $b$ & loss 
			& $b$ & loss \\ \hline
			49 & 0.1 	& 40 & 2.5 & 90 & 0.1 & 33 &  9.0 & 85 & 4.8 & 90 & 14.2 & 39 & 7.9  \\
			14 & 18.5 &    &     & 55 & 1.5 & 13 & 28.7 & 60 & 0.9 & 52 &  1.8 &  &  \\
			13 & 0.3 	&    &     &    &     &  5 &  7.6 & 41 & 6.3 & 33 &  0.8 &  &  \\
			11 & 1.0 	&    &     &    &     &    &      &    &     & 23 & 24.5 &  &  \\
			   &      &    &     &    &     &    &      &    &     & 19 &  0.5 &  &  \\ \hline
		\end{tabular}
		\caption{Calibration for Goodwill as sets of binary cash-or-nothing American options.  When the barrier $b$, as a percentage of mid 2007 stock price, is hit then the Goodwill loss (percent) occurs (see text for details). \label{t:GWcalibration}}
\end{table}

\FloatBarrier
\section{Results}

\subsection{CONSTANT and PROGRESSIVE models}

We calibrate the PROGRESSIVE model using historical data mid-2007 to end-2011 on seven US banks see Figure \ref{f:EQandGW} (right panel) and Table \ref{t:GWcalibration}.  Stock implied volatility is from Bloomberg using ATM volatility at the longest consistently available quote (18 months).  We use the 5Y CDS spread as representative together with the 5Y swap rate for discounting.  The Goodwill that was not lost over this period we assign to the CONSTANT model.

\begin{figure}[htbp]
	\centering
		\includegraphics[width=\linewidth]{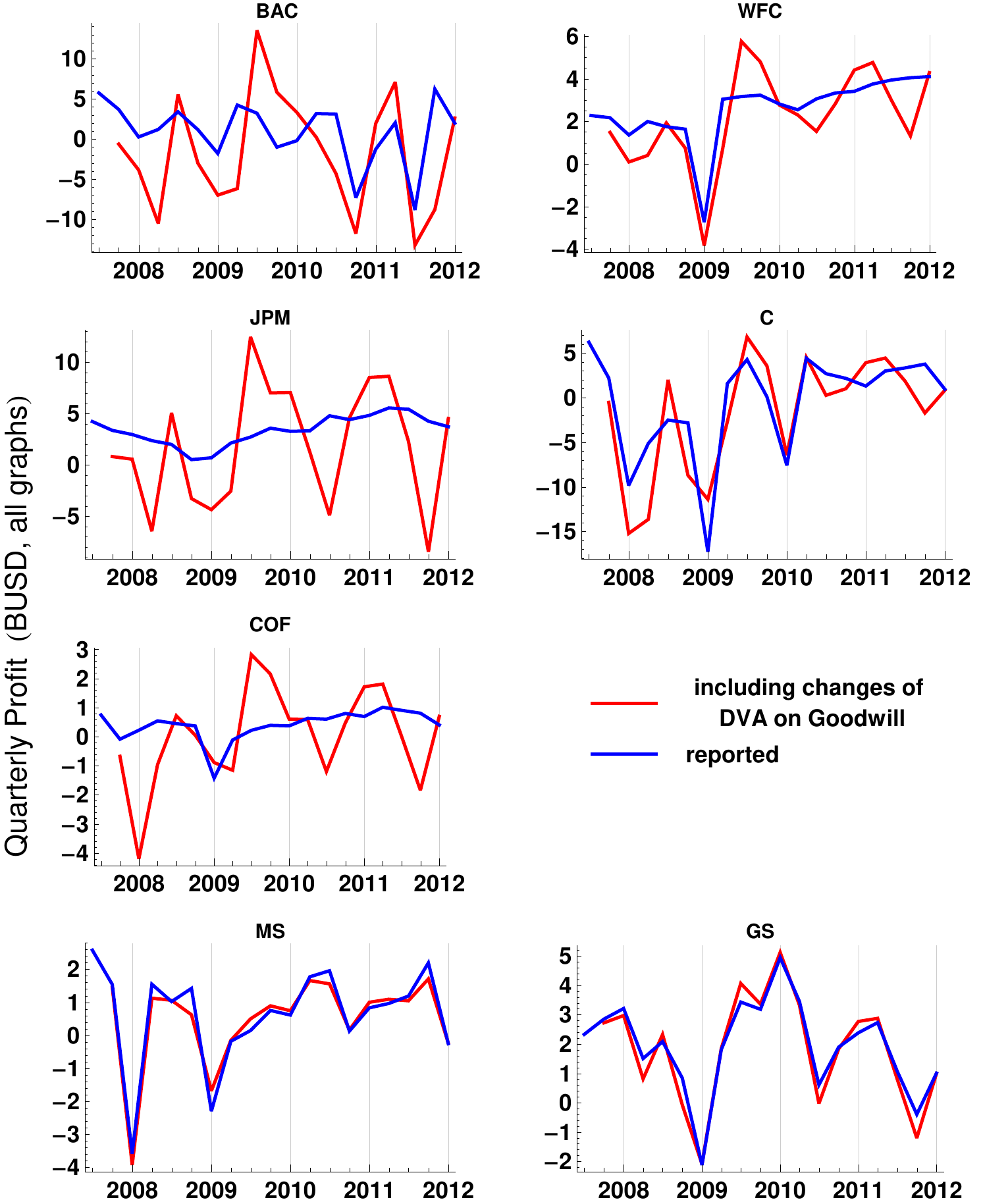}
	\caption{Effect on reported quarterly profits of including changes in DVA on Goodwill.  Five large banks are shown as well as two investment banks (bottom row).  DVA on Goodwill model is the combined PROGRESSIVE+CONSTANT using historically calibrated barriers.}
	\label{f:GWprofit}
\end{figure}

Figure \ref{f:GWprofit} shows the effect on reported quarterly profits of including changes in DVA on Goodwill.  The two investment banks (GS and MS) see little effect on their reported profits.  This is because they have little Goodwill relative to their profits and it was little affected by writedowns over the crisis period.  One major bank (C) showed initially large effects which were later much reduced.  This is because C wrote off a significant fraction of Goodwill over the crisis and subsequently was only affected by CDS changes.  The remaining four major banks show volatile effects over the crisis period.  This reflects their large amounts of Goodwill, the crisis, and in some cases changes resulting from acquisitions.

\subsection{AMORTIZING model}

\begin{figure}[htbp]
	\centering
		\includegraphics[width=\linewidth]{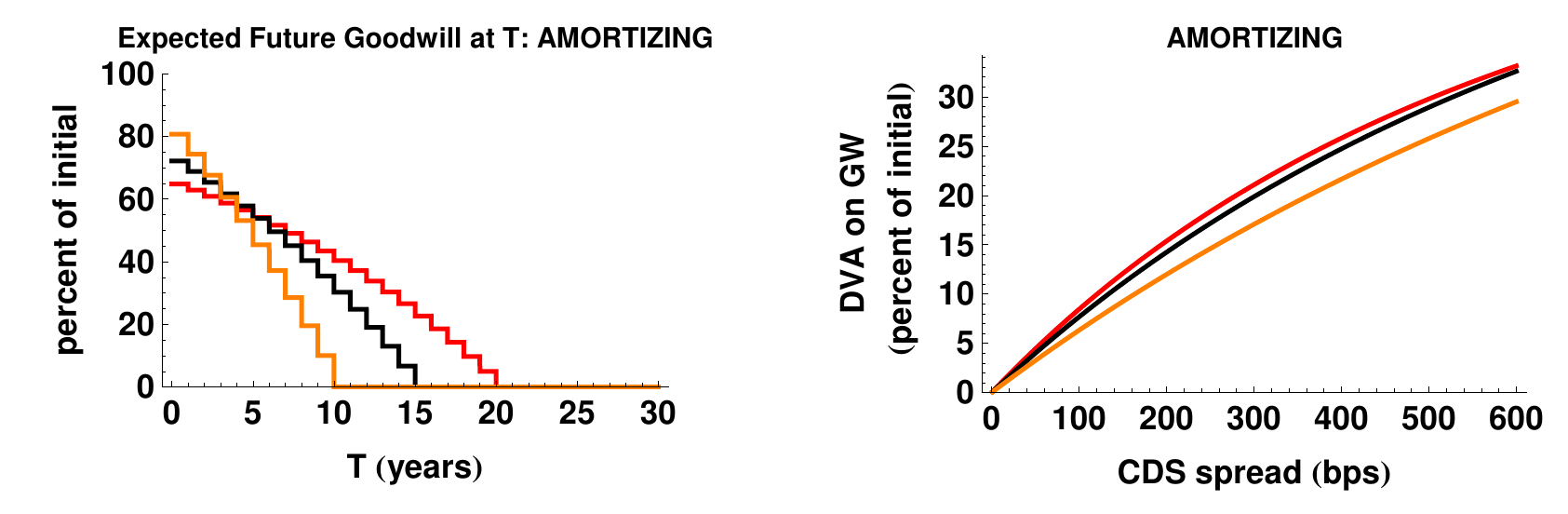}
	\caption{{\bf Left panel} Expected future values of potential tax credits from Goodwill as percentage of initial value under AMORTIZING model.  Straight line amortization over $n_A=10,\ 15,\ 20$ years is displayed.  {\bf Right panel} shows the DVA on these expected future tax credits from Goodwill amortization for a range of CDS spreads, with hazard rates calculated assuming 40\%\ recovery, and flat 5\%\ discount curve.}
	\label{f:GWamort}
\end{figure}

Figure \ref{f:GWamort},  left panel, shows the value of future Goodwill for the AMORTIZING model, which is inspired by tax considerations.    The the staircase effect from paying tax yearly is evident.  There is a strong dependence on the length of the amortization period (which can be set by law).

The DVA on tax credits from Goodwill amortization is shown in Figure \ref{f:GWamort} right panel. There is a relatively small range of values for a wide range of amortizing lengths, 10--20 years.  The amortization range is important in that it shows that the results are robust against deferrement of the use of the tax credits.  This potentially captures the case where the credits can only be used one third to half the time.  If, net, there are no profits over a long continuous period then the tax credits may not be used (also because there will be other sources of tax credits).  However, such cases are likely already captured by the default probability.

\section{Discussion}

DVA on liabilities and derivatives is well established \cite{Burgard2011a,Brigo2011a,
Cesari2010a}, even to the extent or proscriptive accounting rules in some jurisdictions, e.g. FAS157.  DVA on assets, as far as the authors are aware, has had only limited attention \cite{Kenyon2012a}.  We have presented a concrete example of an asset whose value is dependent on the default of its owner (Goodwill) and shown how the potential value lost on default can be hedged using an extension of \cite{Burgard2011a}.  Calibrating our models to seven US banks over the crisis we have shown that the effect of changes DVA on assets can have significant effects on reported profits.

This work complements existing studies on valuation adjustments to do with creditworthiness, collateral, and funding \cite{Burgard2011a, Brigo2011a, 
Cesari2010a, Brigo2011d, Pallavicini2011b, 
Crepey2012b, Brigo2012a}.
We note that DVA is specifically excluded from regulatory capital \cite{BCBS-DVA} but is no less a trading reality.  DVA hedging by proxy has been suggested (WSJ).  This works for spread changes, but not default events: for example, imagine if Morgan Stanley had used CDS on Lehman.

Technically, as in \cite{Burgard2011a} and pointed out by \cite{Kenyon2012a} (Sections 8.3.2 and 8.3.3), the key to analytic tractability is the use of repo accounts for financing.  Without this, analytic tractability is limited.

As well as the accounting point of view of our CONSTANT and PROGRESSIVE models we consider potential losses, and their hedging, relating to a tax point of view with our AMORTIZING model.  Thus we contribute to widening the debate over the scope and application of DVA.    

Banks have reported large changes in profits from effects of their own creditworthiness on liabilities.  Our investigation suggests that including credititworthiness on assets can make the picture even more volatile.  In balance sheet terms, and hedging terms, this volatility is real.  

\section*{Acknowledgments}

The authors would like to acknowledge stimulating discussions with David Southern.  Of course, all errors and opinions remain their own.

\bibliographystyle{alpha}
\bibliography{kenyon_general}

\newpage
\section*{Appendix (online only): Market Data}

All data is from Bloomberg.  Bank data has been aligned to common quarter ends.

\be
\begin{array}{cccccccc}
 \text{Goodwill (MUSD)} & \text{BAC} & \text{WFC} & \text{JPM} & \text{C} & \text{COF} & \text{MS} & \text{GS} \\ \hline
 \text{30Jun07} & 65845. & 11983. & 45254. & 39231. & 13612. & 2977. & 3145. \\
 \text{30Sep07} & 67433. & 12018. & 45335. & 39949. & 12952.8 & 2554. & 3161. \\
 \text{31Dec07} & 77530. & 13106. & 45270. & 41053. & 12830.7 & 3024. & 3321. \\
 \text{31Mar08} & 77872. & 13148. & 45695. & 43622. & 12826.4 & 3053. & 3507. \\
 \text{30Jun08} & 77760. & 13191. & 45993. & 42386. & 12826.7 & 2988. & 3530. \\
 \text{30Sep08} & 81756. & 13520. & 46121. & 39662. & 12815.6 & 2961. & 3553. \\
 \text{31Dec08} & 81934. & 22627. & 48027. & 27132. & 11964.5 & 2243. & 3523. \\
 \text{31Mar09} & 86910. & 23825. & 48201. & 26410. & 13076.8 & 2226. & 3528. \\
 \text{30Jun09} & 86246. & 24619. & 48288. & 25578. & 13381.1 & 6836. & 3536. \\
 \text{30Sep09} & 86009. & 24052. & 48334. & 25423. & 13525. & 6977. & 3546. \\
 \text{31Dec09} & 86314. & 24812. & 48357. & 25392. & 13596. & 7162. & 3543. \\
 \text{31Mar10} & 86305. & 24819. & 48359. & 25662. & 13589.3 & 7169. & 3575. \\
 \text{30Jun10} & 85801. & 24820. & 48320. & 25201. & 13588. & 6749. & 3548. \\
 \text{30Sep10} & 75602. & 24831. & 48736. & 25797. & 13593. & 6766. & 3507. \\
 \text{31Dec10} & 73861. & 24770. & 48854. & 26152. & 13591. & 6739. & 3495. \\
 \text{31Mar11} & 73869. & 24777. & 48856. & 26339. & 13597. & 6743. & 3322. \\
 \text{30Jun11} & 71074. & 24776. & 48882. & 26621. & 13596. & 6744. & 3323. \\
 \text{30Sep11} & 70832. & 25038. & 48180. & 25496. & 13593. & 6709. & 3643. \\
 \text{31Dec11} & 69967. & 25115. & 48188. & 25413. & 13592. & 6686. & 3802. \\ \hline
\end{array}
\ee
\be
\begin{array}{cccccccc}
 \text{Stock Prices (USD)} & \text{BAC} & \text{WFC} & \text{JPM} & \text{C} & \text{COF} & \text{MS} & \text{GS} \\ \hline
 \text{30Jun07} & 48.89 & 35.17 & 48.45 & 512.9 & 78.44 & 69.63 & 216.75 \\
 \text{30Sep07} & 50.27 & 35.62 & 45.82 & 466.7 & 66.43 & 63. & 216.74 \\
 \text{31Dec07} & 41.26 & 30.19 & 43.65 & 294.4 & 47.26 & 53.11 & 215.05 \\
 \text{31Mar08} & 37.91 & 29.1 & 42.95 & 214.2 & 49.22 & 45.7 & 165.39 \\
 \text{30Jun08} & 23.87 & 23.75 & 34.31 & 167.6 & 38.01 & 36.07 & 174.9 \\
 \text{30Sep08} & 35. & 37.53 & 46.7 & 205.1 & 51. & 23. & 128. \\
 \text{31Dec08} & 14.08 & 29.48 & 31.53 & 67.1 & 31.89 & 16.04 & 84.39 \\
 \text{31Mar09} & 6.82 & 14.24 & 26.58 & 25.3 & 12.24 & 22.77 & 106.02 \\
 \text{30Jun09} & 13.2 & 24.26 & 34.11 & 29.7 & 21.88 & 28.51 & 147.44 \\
 \text{30Sep09} & 16.92 & 28.18 & 43.82 & 48.4 & 35.73 & 30.88 & 184.35 \\
 \text{31Dec09} & 15.06 & 26.99 & 41.67 & 33.1 & 38.34 & 29.6 & 168.84 \\
 \text{31Mar10} & 17.85 & 31.12 & 44.75 & 40.5 & 41.41 & 29.29 & 170.63 \\
 \text{30Jun10} & 14.37 & 25.6 & 36.61 & 37.6 & 40.3 & 23.21 & 131.27 \\
 \text{30Sep10} & 13.1025 & 25.115 & 38.06 & 39.1 & 39.55 & 24.68 & 144.58 \\
 \text{31Dec10} & 13.34 & 30.99 & 42.42 & 47.3 & 42.56 & 27.21 & 168.16 \\
 \text{31Mar11} & 13.33 & 31.71 & 46.1 & 44.2 & 51.96 & 27.32 & 158.6 \\
 \text{30Jun11} & 10.96 & 28.06 & 40.94 & 41.64 & 51.67 & 23.01 & 133.09 \\
 \text{30Sep11} & 6.12 & 24.12 & 30.12 & 25.615 & 39.63 & 13.51 & 94.55 \\
 \text{31Dec11} & 5.56 & 27.56 & 33.25 & 26.31 & 42.29 & 15.13 & 90.43 \\ \hline
\end{array}
\ee
\be
\begin{array}{cccccccc}
 \text{Profit (MUSD)} & \text{BAC} & \text{WFC} & \text{JPM} & \text{C} & \text{COF} & \text{MS} & \text{GS} \\ \hline
 \text{30Jun07} & 5761. & 2279. & 4234. & 6226. & 750.372 & 2582. & 2333. \\
 \text{30Sep07} & 3698. & 2173. & 3373. & 2212. & -81.658 & 1543. & 2854. \\
 \text{31Dec07} & 268. & 1361. & 2971. & -9833. & 226.568 & -3588. & 3215. \\
 \text{31Mar08} & 1210. & 1999. & 2373. & -5111. & 548.504 & 1551. & 1511. \\
 \text{30Jun08} & 3410. & 1753. & 2003. & -2495. & 452.905 & 1026. & 2087. \\
 \text{30Sep08} & 1177. & 1637. & 527. & -2815. & 374.139 & 1425. & 845. \\
 \text{31Dec08} & -1789. & -2734. & 702. & -17263. & -1421.55 & -2295. & -2121. \\
 \text{31Mar09} & 4247. & 3045. & 2141. & 1593. & -108.062 & -177. & 1814. \\
 \text{30Jun09} & 3224. & 3172. & 2721. & 4279. & 223. & 149. & 3435. \\
 \text{30Sep09} & -1001. & 3235. & 3588. & 101. & 394. & 757. & 3188. \\
 \text{31Dec09} & -194. & 2823. & 3278. & -7579. & 376. & 617. & 4948. \\
 \text{31Mar10} & 3182. & 2547. & 3326. & 4428. & 636.263 & 1776. & 3456. \\
 \text{30Jun10} & 3123. & 3062. & 4795. & 2697. & 608. & 1960. & 613. \\
 \text{30Sep10} & -7299. & 3339. & 4418. & 2168. & 803. & 131. & 1898. \\
 \text{31Dec10} & -1244. & 3414. & 4831. & 1309. & 697. & 836. & 2387. \\
 \text{31Mar11} & 2049. & 3759. & 5555. & 2999. & 1016. & 968. & 2735. \\
 \text{30Jun11} & -8826. & 3948. & 5431. & 3341. & 911. & 1193. & 1087. \\
 \text{30Sep11} & 6232. & 4055. & 4262. & 3771. & 813. & 2199. & -393. \\
 \text{31Dec11} & 1991. & 4107. & 3728. & 956. & 407. & -250. & 1013. \\ \hline
\end{array}
\ee
\be
\begin{array}{cc}
 \text{} & \text{USD 5Y swap rate (\%)} \\ \hline
 \text{30Jun07} & 5.5095 \\
 \text{30Sep07} & 4.8115 \\
 \text{31Dec07} & 4.1865 \\
 \text{31Mar08} & 3.3105 \\
 \text{30Jun08} & 4.2715 \\
 \text{30Sep08} & 4.0223 \\
 \text{31Dec08} & 2.135 \\
 \text{31Mar09} & 2.2298 \\
 \text{30Jun09} & 2.94 \\
 \text{30Sep09} & 2.6485 \\
 \text{31Dec09} & 2.978 \\
 \text{31Mar10} & 2.7275 \\
 \text{30Jun10} & 2.0698 \\
 \text{30Sep10} & 1.538 \\
 \text{31Dec10} & 2.182 \\
 \text{31Mar11} & 2.402 \\
 \text{30Jun11} & 1.998 \\
 \text{30Sep11} & 1.2594 \\
 \text{31Dec11} & 1.233 \\ \hline
\end{array}
\ee
\be
\begin{array}{cccccccc}
 \text{5Y CDS spread (bps)} & \text{BAC} & \text{WFC} & \text{JPM} & \text{C} & \text{COF} & \text{MS} & \text{GS} \\ \hline
 \text{30Jun07} & 12.5 & 10.5 & 19.33 & 11.667 & 27. & 33.813 & 34.938 \\
 \text{30Sep07} & 33.17 & 27.771 & 36.002 & 33.406 & 41.364 & 55.244 & 45.096 \\
 \text{31Dec07} & 48.868 & 59.67 & 49.231 & 71.396 & 214.15 & 97.201 & 66.33 \\
 \text{31Mar08} & 107.083 & 95.63 & 110.9 & 182.055 & 278.911 & 172.339 & 148.337 \\
 \text{30Jun08} & 110.82 & 113.82 & 104.066 & 139.363 & 316.725 & 184.575 & 136.005 \\
 \text{30Sep08} & 151.398 & 150.916 & 143.911 & 305.496 & 344.736 & 825.094 & 419.389 \\
 \text{31Dec08} & 120.617 & 118.985 & 121.318 & 193.053 & 189.093 & 413.111 & 290.132 \\
 \text{31Mar09} & 395.3 & 297.775 & 201.125 & 631.526 & 288.941 & 399.749 & 285.964 \\
 \text{30Jun09} & 221.858 & 155.311 & 108.823 & 425.131 & 153.733 & 211.698 & 151.192 \\
 \text{30Sep09} & 120.475 & 79.674 & 69.592 & 190.186 & 74.02 & 138.544 & 106.465 \\
 \text{31Dec09} & 101.584 & 91.226 & 50.315 & 172.561 & 75.93 & 114.037 & 91.848 \\
 \text{31Mar10} & 117.406 & 91.519 & 59.265 & 155.575 & 70.806 & 136.483 & 101.694 \\
 \text{30Jun10} & 156.967 & 119.107 & 116.294 & 186.84 & 102.699 & 267.296 & 189.07 \\
 \text{30Sep10} & 165.757 & 105.214 & 85.098 & 174.911 & 85.116 & 182.203 & 153.283 \\
 \text{31Dec10} & 180.764 & 105.861 & 86.162 & 148.86 & 85.226 & 170.722 & 125.94 \\
 \text{31Mar11} & 135.527 & 81.227 & 70.215 & 124.922 & 70.196 & 139.859 & 113.433 \\
 \text{30Jun11} & 156.763 & 95.093 & 79.283 & 136.756 & 80.388 & 161.813 & 136.673 \\
 \text{30Sep11} & 422.085 & 158.126 & 163.274 & 282.172 & 125.019 & 466.757 & 324.692 \\
 \text{31Dec11} & 411.59 & 144.112 & 147.216 & 285.49 & 110.243 & 419.927 & 325.2 \\ \hline
\end{array}
\ee
\be
\begin{array}{cccccccc}
 \text{ATM Implied Volatility (\%, 18M)} & \text{BAC} & \text{WFC} & \text{JPM} & \text{C} & \text{COF} & \text{MS} & \text{GS} \\ \hline
 \text{30Jun07} & 20.153 & 21.114 & 23.465 & 22.164 & 26.715 & 30.141 & 29.179 \\
 \text{30Sep07} & 22.15 & 25.895 & 27.07 & 25.365 & 34.095 & 30.141 & 30.352 \\
 \text{31Dec07} & 29.806 & 31.803 & 31.327 & 35.599 & 48.298 & 37.503 & 37.879 \\
 \text{31Mar08} & 38.975 & 41.903 & 42.64 & 49.631 & 56.769 & 48.111 & 43.21 \\
 \text{30Jun08} & 46.853 & 42.207 & 44.711 & 49.798 & 60.585 & 46.201 & 40.154 \\
 \text{30Sep08} & 48.197 & 49.204 & 40.478 & 50.384 & 68.035 & 82.419 & 49.466 \\
 \text{31Dec08} & 72.616 & 61.777 & 60.783 & 81.9 & 84.179 & 81.272 & 59.742 \\
 \text{31Mar09} & 118.159 & 92.138 & 77.834 & 137.187 & 97.422 & 85.786 & 66.599 \\
 \text{30Jun09} & 60.789 & 56.672 & 49.282 & 69.778 & 63.009 & 46.964 & 41.714 \\
 \text{30Sep09} & 58.554 & 48.07 & 42.539 & 66.669 & 56.443 & 47.55 & 35.94 \\
 \text{31Dec09} & 42.142 & 40.008 & 35.169 & 53.108 & 42.681 & 37.275 & 33.424 \\
 \text{31Mar10} & 35.517 & 31.031 & 30.693 & 43.556 & 38.451 & 35.783 & 29.512 \\
 \text{30Jun10} & 47.84 & 44.015 & 41.516 & 53.491 & 47.969 & 45.23 & 40.54 \\
 \text{30Sep10} & 44.228 & 39.699 & 40.187 & 49.395 & 46.114 & 39.791 & 33.68 \\
 \text{31Dec10} & 39.834 & 35.219 & 31.94 & 38.219 & 36.853 & 33.96 & 29.122 \\
 \text{31Mar11} & 34.974 & 30.776 & 27.26 & 30.471 & 32.413 & 32.015 & 25.691 \\
 \text{30Jun11} & 35.876 & 31.059 & 28.746 & 31.43 & 31.445 & 34.133 & 28.538 \\
 \text{30Sep11} & 37.873 & 45.547 & 27.6 & 32.499 & 31.532 & 66.342 & 46.45 \\
 \text{31Dec11} & 57.723 & 37.03 & 39.868 & 49.602 & 39.906 & 57.681 & 41.55 \\ \hline
\end{array}
\ee

\end{document}